\documentstyle[12pt]{article}

\renewcommand{\a}{\alpha}
\renewcommand{\b}{\beta}

\renewcommand{\d}{\delta}
\newcommand{\D}{\Delta} 

\newcommand{\ep}{\epsilon}

\newcommand{\la}{\lambda}

\newcommand{\si}{\sigma}
\newcommand{\Si}{\Sigma}
\newcommand{\th}{\theta}

\def\chib{{\bar \chi}}

\def\psib{{\bar \psi}}

\def\thb{{\bar \theta}}

\def\ad{{\dot \a}}

\def\psidag{{\psi^{\dagger}}}
\def\Xd{{\dot X}}

\def\tr{{\rm tr}}

\newcommand{\mC}{{{\rm C}\hspace*{-0.9ex}\rule{0.05ex}%
       {1.4ex}\hspace*{0.9ex}}}

\def\llap#1{\hbox to 0pt{\hss#1}}
\def\pola{a\llap{\hbox{\char'30\kern-1.2pt}}}
\def\pole{e\llap{\hbox{\char'30\kern-.8pt}}}
\newcommand{\p}{\partial}

\newcommand{\non}{\nonumber\\}

\def\half{{\mbox{\small  $\frac{1}{2}$}}}
\def\ovsqrt{\frac{1}{\sqrt{2}}}

\newcommand{\beq}{\begin{equation}}
\newcommand{\eeq}{\end{equation}}
\newcommand{\beqa}{\begin{eqnarray}}
\newcommand{\eeqa}{\end{eqnarray}}
\newcommand{\ben}{\begin{enumerate}}
\newcommand{\een}{\end{enumerate}}
\newcommand{\bit}{\begin{itemize}}
\newcommand{\eit}{\end{itemize}}

\newcommand{\refeq}[1]{(\ref{#1})}

\newcommand{\prl}[1]{{ \it Phys.~Rev.~Lett.}~{\bf {#1}}}
\newcommand{\plb}[1]{{ \it Phys.~Lett.}~{\bf {#1B}}}
\newcommand{\pla}[1]{{ \it Phys.~Lett.}~{\bf {#1A}}}
\newcommand{\npb}[1]{{ \it Nucl.~Phys.}~{\bf B{#1}}}

\newcommand{\cmp}[1]{{\it Comm.~Math.~Phys}~{\bf {#1}}}

\newcommand{\jgp}[1]{{\it  J.~ Geom.~Phys.}~{\bf {#1}}}

\def\Xh{{\hat X}}
\def\Xd{{\dot X}}
\def\cF{{\cal F}}
\def\cN{{\cal N}}
\def\ninfty{N$_c\to\infty$ }

\begin{document}

\baselineskip 19pt plus 2pt
\begin{titlepage}
\renewcommand{\thefootnote}{\fnsymbol{footnote}}
\begin{flushright}
\parbox{1.5in}
{ LMU-HEP-98-04 \\
hep-th/9802182\\}
\end{flushright}
\vspace*{.5in}
\begin{centering}
{\Large 
Matrix Model for Yang-Mills Interactions
}\\
\vspace{2cm} 
{\large        
Jacek Pawe\l czyk\footnote{On leave from Institute of
    Theoretical Physics, Warsaw University, Ho\.{z}a 69, PL-00-681 Warsaw,
    Poland. Email: Jacek.Pawelczyk@fuw.edu.pl}
}\\
\vspace{.5cm}
        {\sl 
Sektion Physik, Universit\"at M\"unchen\\
Theresienstrasse 37
80333 M\"unchen, Germany}\\
\vspace{.5in}
\end{centering}
\begin{abstract}
We introduce a $N_c\times N_c$ matrix model with $\cN=2$ supersymmetries 
and show its relation to the topological rigid string and the topological
YM$_2$.  This allows to connect the latter two theories directly. Moreover the
construction leads to a new insight in the \ninfty limit.
Finally a quantum mechanical matrix
theory is proposed which may describe light-cone (light-front) dynamics of gauge fields.
\end{abstract}
\vfill
\end{titlepage}

\renewcommand{\thefootnote}{\arabic{footnote}}
\setcounter{section}{0}
\setcounter{footnote}{0}

It is believed that the dynamics of Yang-Mills fields should be given by a string theory.
Unfortunately no appropriate string theory has been built yet. There are several reasons to believe that
the correct theory will incorporate rigidity \cite{polrig,my:inst}. 
But the rigid
string can not be quantized in standard  manner although its topological sector
is amazingly simple \cite{my:holo}.  

In this paper we shall consider quantum mechanical matrix models meant to describe
dynamics of the Yang-Mills fields. Thus they are not supersymmetric in
space-time.  
Quantum mechanics of
free particles of any spin $s$ is well understood \cite{town}.  It is given by
$s=\cN/2$ extended 1d supergravity coupled to matter multiplets.  
One would like to have a similar
picture for many body interacting systems e.g.  gauge particles. 
The hope is that matter multiplets in form of  $N_c\times N_c$ matrices
will provide a convenient representation of multi-particle states and will allow to introduce 
interaction between those states.  The necessary gauge symmetry must be
introduced on the world-line. 
Unfortunately this construction is not known if  the world-line SUGRA  is dynamical.
There are two ways out. One can limit the discussion to theories which are world-line SUGRA
independent i.e. topological. As we deal with supersymmetric quantum mechanics,
the proposed matrix models posses
natural topological symmetry if one assumes that the matrices do not depend on
the world-line time variable.
 
An alternative way is to construct a non-relativistic theory with globally
well defined world-line time $\tau$ for a multi-particle system. 
This theory maybe interpreted as a covariant theory in the light-cone gauge (or
on the light-front) for
which  $\tau\propto x^+$. The advantages of these frames in investigations of
many-particle systems are well known \cite{ks,glazek}.
For the light-front frame one hopes to recover
Lorentz invariance in the \ninfty limit.  There is also a heuristic need for infinite matrices.
Because we want to think about the matrix model as quantum mechanics of gauge particles
we should not expect to get gauge invariant states built from a finite number of
gluons. 

This work was also inspired by the recent progress in (unifying)
M/string theory.  According to \cite{matrix} M-theory in the light-front frame
is given by a certain space-time supersymmetric, quantum mechanical matrix model. Its basic
ingredients are D0-branes. These are the partons.  

The outline of the paper is the following. In the first section we shall
recall certain facts about $\cN=2$ world-line supergravity coupled to matter and we shall show
how it is related to abelian gauge particles. This section has a review character as
all the results are well known. The next section is devoted to 
(time-independent) topological matrix models and their relation to 2d
topological strings
and to the 2d topological YM$_2$ theory of Witten \cite{witten:ym2}. In the last
section we shall shortly speculate on a dynamical matrix model. 

\section{Spin 1 ($\cN=2$) particle}
\label{sec:particle}

It is well known that $\cN=2$ SUGRA  on the world-line coupled to matter fields
describes spin 1 particles  
in the target space time \cite{town}. The relevant action is
\beq
S=\half\int d\tau
\left(\frac{1}{e}(\Xd-i(\chi\psib+\chib\psi))^2+i\psi D\psib+i\psib D\psi
+eF^2\right) 
\label{g:model}
\eeq
where $\psi=\ovsqrt(\psi_1+i\psi_2)$ and $\psi_{1,2}$ are real.  
Also $D\psi\equiv(\p_\tau-if)\psi$. The same convention
holds for $\ep$ and $d\chi$.  In \refeq{g:model} $(e,\chi,\chib,f)$ is the gravity
multiplet and
$(X,\psi,\psib,F)$ is the matter multiplet carrying a space-time
index\footnote{Wherever it will be possible this index 
will be suppressed in order to simplify notation.}. 

The local supersymmetry transformations are
\beqa
\d X&=&i\ep\psib+h.c.\non
\d\psi&=&\frac{\ep}{e}(-\Xd+i(\chi\psib+\chib\psi)+ieF)\non
\d F&=&\frac{\ep}{e}
(D\psib+i\chi F+\frac{1}{e}\chib\Xd+\frac{i}{e}\chi\chib\psib)+h.c.\\
\d e&=&-2i\ep+h.c.\non
\d\chi&=&{D\ep}\non
\d f&=&0\nonumber
\label{sugra}
\eeqa
The theory possesses a local U$_f$(1) symmetry of the 2 supersymmetry charges for
which $f$ is the gauge field.
The constraints on physical states are $Q_i|phys>= H|phys>=T|phys>=0$ $(i=1,2)$
where $Q,\; H,\; T^0$ are the  generators of the supersymmetry, the  local
reparameterizations and  the U$_f$(1), respectively. 
The standard quantization procedure gives 
\beq
\{\psi_i^m,\psi_j^n\}=\d_{ij}\eta^{mn}.
\eeq
In this paper we shall mainly deal with $T^0$ thus we write it 
explicitly,
\beq
T^0=i\psi_1^m\psi_2^m.
\label{uone}
\eeq
For 4d Minkowski space-time the constraint $T|phys>=0$
implies that physical states are in
the symmetric part of the  $(\half,\half)$ representation
of SL(2,$\mC$). So we write $|phys>=\Psi_{\a\b}$.
Then the equation $Q|phys>=0$ reads
\beq 
\p_\ad^{\;\a}\Psi_{\a\b}=0
\label{eq:state}
\eeq
We can write the state $\Psi_{\a\b}$ as $\Psi_{\a\b}=
\si^{mn}_{\a\b}\cF_{mn}$, where $\cF$ is an antisymmetric tensor. 
Because $*\si^{mn}=i\si^{mn}$ we get $\cF_{mn}=F_{mn}-iF_{mn}$ for real $F_{mn}$.
Then \refeq{eq:state} yields the Maxwell equations,
\beq
\p^m F_{mn}=0,\quad \p^m *F_{mn}=0.
\eeq

\section{Variables and interactions}
The purpose of this section is to generalize the first quantized version  of  a
single gauge  particle to the multi-particle case. 
In the following we shall keep
the U$_f$(1) gauge field $f$ and global $\cN=2$ SUSY as the only remnant of
the local SUSY symmetry \refeq{sugra} of the theory. 
We shall in sec.\ref{sec:dmod} that the U$_f$(1) is necessary
for having proper number of degrees  of freedom.  We
introduce matter fields $(X,\psi,F)$ as hermitian 
$N_c\times N_c$ matrices and put all matter in
a single $\cN=2$ supermultiplet
\beq
\Xh=X+i(\th\psib+\thb\psi)+\th\thb F
\eeq
We also impose a SU(N$_c$)
local gauge symmetry on the world-line under which the matter fields
$(X,\psi,F)$ will transform in the adjoint
representation of SU(N$_c$). Thus we also need a gauge field $A$.
The SUSY transformation rules are
\beqa
\d X&=&i\ep\psib+h.c.\\
\d\psi&=&{\ep}(-DX+ieF)\\
\d F&=&{\ep}D\psib+h.c.
\label{susym}
\eeqa
All derivatives in the above are defined with the SU(N$_c$) connection $A$ i.e.
$DX\equiv(\p_\tau X-i[A,X])$ and $D\psib\equiv (\p_\tau \psib+if\psib-i[A,\psib])$.
The kinetic term is given by
\beqa
L_0&=&\tr(|D_\th \Xh|^2)|_{\th\thb}\non
&=&\tr\left((DX)^2+i\psi D\psib+i\psib D\psi+F^2\right) 
\label{kin}
\eeqa
where $D_\th \Xh\equiv\p_\th \Xh-i\thb D\Xh$. The covariant derivative is
understood as above i.e. it has an extra contribution when acting on $\psi$ and $\psib$.
Now we build the interaction of  our system. We want the
interaction to preserve shift symmetry  $X\to X+a{\bf 1}$.
This corresponds to ordinary shift symmetry of space time coordinates. 
If we bound considerations to terms which exist in
space-times of any dimension  then 
it appears that there is a unique lowest order non-derivative operator: 
\beqa 
\la\tr([\Xh^m,\Xh^n]^2)|_{\th\thb}&=&
4\la \tr([X^m,X^n][X^m,F^n]-[X^m,X^n]\{\psi^m,\psib^n\}\non
&&\hskip1cm-[X^m,\psi^n][X^m,\psib^n]-[X^m,\psi^n][\psib^m,X^n])\label{int}
\eeqa
In various specific dimensions we have more possibilities.  For 3d space we could have 
\beqa
\ep_{mnr}\tr([\Xh^m,\Xh^n]\Xh^r)|_{\th\thb}\nonumber
\eeqa
while 4d spaces allow for:
\beqa
\la\int d\tau \tr([\Xh^m,\Xh^n]*[\Xh^m,\Xh^n])|_{\th\thb}\nonumber
\eeqa
where * denotes the Hodge star in the target space. If one extends the gauge multiplet then there are
more choices.  We are going to discuss one of such operators in
section \ref{sec:toptwo}. One can also add the unity operator $\tr({\bf 1})=N$ 
to the action. It looks trivial but  plays an
important role in the limit of infinite matrices.  Terms with derivatives
are also allowed but they will not be discussed here.

\section{Topological matrix models}
In this section we shall construct several topological matrix models and show their equivalence with
various 2d topological theories.  We shall see that the BRST algebra is intimately related to 
$\cN=2$ quantum mechanics of the previous section. 
Relying on this relation we shall  show that the 
topological rigid string \cite{horava} has a natural matrix counterpart. As a bonus we shall 
get immediately an extra BRST-like charge \cite{horava}. 
After taking the \ninfty limit we shall obtain the topological rigid string.
A similar procedure will be applied to  the matrix topological YM$_2$.
After compactification of the target space on a torus we shall also get
topological YM$_2$ \cite{witten:ym2}.  Moreover we shall claim that both theories
are equivalent. This will lead to a direct comparison of the topological rigid
string and topological YM$_2$.
Although all calculations are made for flat spaces we believe that the results
should hold for arbitrary Riemann surfaces. The reason is that all the
theories have the same topological symmetry and define the same moduli problem. 
This is apparent in the matrix formulation.  
We also notice that the trace part of all matrices will not play any role in this
section.

We take all quantities to be 1d time independent. 
Consequently we suppress also the 1d gauge fields form the 
algebra. Then \refeq{susym} is
\beqa
\d X&=&i(\ep_1\psi_1+\ep_2\psi_2)\non
\d\psi_1&=&-\ep_2F\non
\d\psi_2&=&\ep_1F\label{brst}\\
\d F&=&0\nonumber
\eeqa
It is clear that we have two candidates for BRST charges $Q_i$ ( $i=1,2$).  They
also respect $\{Q_1,Q_2\}=0$. We choose $Q_1$ to be our BRST charge. Then
$\psi_1$ is the ghost of the topological symmetry $\d X=arbitrary\; matrix$. Hence
($\psi_2,F$) form an anti-ghost system.

\subsection{Topological rigid string}

In this subsection we shall consider the action 
\beq
S=\{Q_1,V\},\quad V=\tr(\psi_2^m([X^n,[X^n,X^m]]+aF^m))
\label{topstring}
\eeq
Let us also notice that analogously to \cite{horava} we have
\beq
V=[Q_2,\tr(-\frac{i}{4}[X^n,X^m]^2+a\, \psi_2^m\psi_1^m]
\eeq
what is a simple consequence of the $\cN=2$ SUSY. 
 For finite
matrices the first term of the  action $S$ is just \refeq{int} while the second is the leftover of
the kinetic term \refeq{kin}.  We can also perturb the theory by a unity
operator. 

The model \refeq{topstring} is localized on matrices respecting
\beq
[X^n,[X^n,X^m]]=0
\label{matharm}
\eeq
For 2d space-times the moduli space of  \refeq{matharm} can be given more explicitly.
Simple calculations (e.g. with help of the Cartan-Weyl basis)
show that \refeq{matharm} is equivalent to
\beq
[X^1,X^2]=0
\label{matflat}
\eeq
We notice that if the target space is a torus, \refeq{matflat} is equivalent to $F_{mn}=0$. 
This will be crucial in establishing an equivalence
of the topological matrix theory \refeq{topstring} and the topological gauge
theory \cite{witten:ym2} for 2d target spaces.

In the following we shall show that a \ninfty limit of \refeq{topstring} leads
naturally to the well known topological rigid string \cite{horava}. 
In this limit we substitute matrices by functions on
a 2d compact parameter space (a Riemann surface $\Si_h$ of genus $h$). 
According to the prescription given in \cite{hoppe} the local SU(N$_c$) symmetry
goes to SDiff($\Si_h$) in this limit and
we substitute
\beq
[A,B]\to \frac{\ep^{ab}\p_a A\p_b B}{\sqrt{g}},\quad a=1,2
\label{sub}
\eeq
where $g$ is the determinant of a metric on the parameter space $\si^a$. 
In this paper we shall choose $g$ to be the induced
metric $g_{ab}=\p_a X\p_b X$.\footnote{One could use another metric on the
  l.h.s. of 
  \refeq{sub} e.g.  a subsidiary elementary metric on $\Si_h$, but this possibility
  will not be discussed here.}
In this way we force the r.h.s. of \refeq{sub}  to be explicitly $X$ dependent no
matter what $A$ and $B$ are.
The choice has several virtues which become apparent during the course of this
article.
After the substitution \refeq{sub} the localization equations \refeq{matharm}
go to 
 \beq 
\D_g X^m=0.  \label{min} 
\eeq 
The Laplacian $\D_g$ is defined with the 2d induced metric $g_{ab}$.  In order
to rewrite the action \refeq{topstring} on $\Si_h$ we substitute
$tr(...)\to\int_{\Si_h}\sqrt{g}(...)$.  With this prescription
\refeq{topstring} defines the topological rigid string \cite{horava} in flat
d-dimensional space-time. We also
notice that $\tr({\bf 1})\to \int_{\Si_h}\sqrt{g}$ i.e. the cosmological
(Nambu-Goto) term.
But what defines $h$? The clue to this point will be obtained in the next
section where we shall discuss the relation of the matrix model with $YM_2$.

\subsection{Topological string  for  2d targets}
\label{sec:toptwo}

In this section we shall show that a slight modification of the previous
construction leads to other  known topological theories in 2d target space-time. 

First we notice that for 2d targets one has to be careful in concluding that 
$\D_g X^m=0\Leftrightarrow t^{mn}=0$ where 
$$
t^{mn}= \frac{\ep^{ab}\p_a X^m\p_bX^n}{\sqrt{g}}
$$
In fact $t^{mn}=0$ has no nontrivial solutions in 2d targets, because $|t^{mn}|=1$.
It is clear that $\D_g X^m=0\Rightarrow t^{mn}=\pm\ep^{mn}$.
Solutions to $t^{mn}=\pm\ep^{mn}$ are given by 
(both signs $\pm$) pseudo-holomorphic curves
\beq
\frac{\ep^{\;b}_a}{\sqrt{g}}\p_b X^m\pm J_n^{\;m}\p_a X^n=0,
\label{ph}
\eeq
where
$J_n^{\;m}$ is 2d complex structure defined by a metric on $M^2$ with
$\ep^{mn}$. Appropriate gluing \cite{cmr,horava} of both spaces of maps
\refeq{ph} gives the space of minimal maps \refeq{min}.  In this sense the
stronger statement $\D_g X^m=0\Leftrightarrow t^{mn}=\pm\ep^{mn}|_{comp}$ holds.
Thus we must conclude
that for 2d targets the substitution \refeq{sub} is, in a sense, renormalized
either to $[X^m,X^n]\to t^{mn}-\ep^{mn}$ or to $[X^m,X^n]\to t^{mn}+\ep^{mn}$.

In 2 dimensions one can build another simple topological theory if one 
extends the gauge multiplet to a full  $\cN=2$ multiplet as in \refeq{brst}. 
Previously under the BRST transformation we had $\d A=0$. Thus we take now 
\beqa
\d \la&=&i(\ep_1\eta_1+\ep_2\eta_2)\non
\d\eta_1&=&-\ep_2A\non
\d\eta_2&=&\ep_1A\label{new}\\
\d A&=&0\nonumber
\eeqa
This is identical to the additional multiplets in section (3.1) of \cite{witten:ym2}.
We take the following gauge fermion
\beq
V= \tr(\eta_2(\ep_{mn}[X^m,X^n]+bA)+[X^m,\la]\psi_1^m)
\label{2d}
\eeq
The \ninfty limit of \refeq{2d} is
\beq
V_\pm=\int_{\Si_h} \eta_2 [\ep_{mn}(\ep^{ab}\p_a X^m\p_bX^n\pm\ep^{mn}\sqrt{g})
+bA\sqrt{g}]+{\ep^{ab}\p_a X^m\p_b\la}\psi_1^m
\label{2dstring}
\eeq
depending on the renormalization prescription discussed above.
Thus \refeq{2d} is localized on both
pseudo-holomorphic curves \refeq{ph} although it seems to have two \ninfty limits \refeq{2dstring}. 
Due to $\D_g X^m=0\Leftrightarrow t^{mn}=\pm\ep^{mn}|_{comp}$ 
we can expect
that theories \refeq{topstring} and \refeq{2d} are equivalent.

\subsection{Topological YM$_2$}

Here we compactify the target space of the theory \refeq{2d}.
In general, the problem is not easy as we know from (M)atrix. 
The compactification on $M^2=T^2$ is the best known 
example \cite{matrix,douglas}. If one does it for the theory \refeq{2d} then one
gets the standard 2d topological YM$_2$ theory \cite{witten:ym2},
\beq
V=\int_{T^2} \left(\phi(\ep^{mn}F_{mn}+bA)+(D_m\la)\psi_1^m\right)
\label{topgauge}
\eeq
with the SU(N$_c$) gauge group. In \refeq{topgauge} $D$ stands for
the standard gauge covariant derivative. We recognize that \refeq{topgauge} is
topological YM$_2$ localized on flat connections $F_{mn}=0$. It would be interesting
to show that a similar correspondence holds for other target spaces e.g. the
Riemann surfaces $M^2=\Si_G$ of  
genus $G$.  We do not have the proof that this is true
but note
that the topological theories, under consideration, 
are insensitive to many changes e.g. they are insensitive to the  metric chosen on  the target manifold. 
Thus we take it for granted that \refeq{2d} 
is the matrix representation of the topological YM$_2$. 

\subsection{Relation between 2d topological string models and topological YM$_2$}

All matrix topological theories with 2d target space described in the previous subsections
define the same moduli problem $[X^m,X^n]=0$. After compactifying space-time on a compact
Riemann surface of genus $G$ this appeared to be equivalent  to $F_{mn}=0$
i.e. the theory of flat SU(N$_c$) connections \cite{witten:ym2}. On the other hand if we first take the
limit \ninfty we get the topological rigid string, or equivalently, the  theory of
pseudo-holomorphic maps $t^{mn}=\pm 1$ from a world-sheet Riemann surface
$\Si_h$ to the 2d target manifold. 
This shows that there should be a relation between both theories.
In fact, the relation is well known \cite{gross,cmr,horava}: both theories are equivalent in  
1/N expansion\footnote{This has been shown for $M^2=\Si_G$ of
genus at least 2 \cite{cmr}. For $G=1$ one needs the cosmological (Nambu-Goto)
term in order the partition function does not vanish. For $G=0$ the situation
is more complicated \cite{sphere}.}.
\beq
Z_{\mbox{YM$_2$}}(SU(N_c),G)=
\sum_h \left(\frac{1}{N_c}\right)^{2h-2}Z_{\mbox{topological strings}}(X:\Si_h\to\Si_G)
\eeq
The above gives the precise meaning to the mysterious genus $h$ which appeared
when we have taken the limit \ninfty. 

\section{Conclusions and outlook}
\label{sec:dmod}

In the previous section we have shown that the matrix model correctly reproduces known
topological theories of Yang-Mills fields. The arguments shed light on the
limit \ninfty and the relation $[A,B]\to \frac{\ep^{ab}\p_a A\p_b
  B}{\sqrt{g}}$. Although the discussed topological models left
several problems 
unsolved we feel obliged to say something about the main initial motivation
of this work i.e. the possible application to the
gauge fields dynamics. In order to do this we must go beyond topological
considerations. 

It is conceivable that the supersymmetric rigid
string \cite{horava} is the proper version of string theory.  On the other hand
this theory is hard to quantize so one can hope that the proposed matrix version will be
easier to deal with. Unfortunately we do not have much to say about it
now. Below we introduce a model which, in a sense, should be simpler because all
variables have ordinary kinetic terms.  

As discussed before we are bound to use  
non-covariant formulation in order to describe multi-particle quantum mechanics. 
Thus we should interpret the theory as the parton
model of gauge particles in the light-front (-cone) gauge \cite{ks}.
The hope is that after taking the \ninfty limit we shall be able to recover
the full Lorentz invariance of the theory. We stick to 4d space-time so the
light-cone target space is 2-dimensional ($m,\,n=1,2$).
Thus the proposed action is
\beq
L=(|D_\th \Xh^m|^2+\la \tr([\Xh^m,\Xh^n]^2))|_{\th\thb}
\label{phys}
\eeq
with the possible addition of a $\tr({\bf 1})$ term. We analyze first the constraints
which follow from the gauge invariance U$_f$(1)$\times $SU(N$_c$). They are
$T^0|phys>=T^a|phys>=0$ ($a=1,...N^2-1$). 
We identify $\psidag_i=\ovsqrt(\psi^1_i+i\psi^2_i)$ and 
$\psi_i=\ovsqrt(\psi^1_i-i\psi^2_i)$ with annihilation and creation
operators. Then the normal ordered operators $T^0,\;T^a$ read
\beq
T^0=i\ep^{ij}\psidag_i^a\psi_j^a,\quad  T^a=f^{abc}(i\psidag_i^b\psi_i^c
+X^{bm}DX^{cm})
\label{con}
\eeq
$T^0$ is the generator of the U$_f$(1) part of the gauge group with the gauge boson
$f$ as  in \refeq{uone},  
while $T^a$ are generators of SU(N$_c$) group with gauge boson $A$. 
In the limit \ninfty the Lagrangian \refeq{phys} with constraints \refeq{con}
is a membrane theory \cite{mem}. 

We recall that in 
sec.\ref{sec:particle} we showed that the world line gauge symmetries leave only
two physical states. The same procedure can be applied for the trace part of
\refeq{phys}. Out of four degenerate states:
\beq
|0>, \psidag_1|0>,\psidag_2|0>,\psidag_1\psidag_2|0>
\label{states}
\eeq
the two in the middle are projected out by the $T^0|phys>=0$ constraint. Hence we get only 
two physical states identified with the polarizations of photon in the light-front
frame.
We believe that similar mechanism holds for the non-abelian part of the model
although we have no firm candidates for gluons.
It is quite
possible that one can not really see them in a simple way i.e. the 
weak coupling Yang-Mills theory might correspond to a non-perturbative region of this matrix
theory \footnote{We refer the
interested reader to the recent work \cite{polch} which makes some observations relevant at
this point.}. 
The full analysis of  \refeq{phys} goes
beyond the scope of this letter and will be the subject of a future publication.
\vskip2cm

{\bf Acknowledgment}.  
I am grateful S.Theisen for many illuminating discussions and reading
the manuscript.
I also thank  K.Gaw{\c e}dzki, S.F\"orste, D.Ghoshal and  J.Madore
for comments and interest in this work.
The research is supported by the Alexander von Humboldt Foundation.


\begin{thebibliography}{99}

\bibitem{polrig} A.M.Polyakov, \npb{268} (1986) 406.
L. Peliti and S. Leibler, \prl{54} (1985) 1690;
D. Foerster, \pla{114} (1986) 115; 
H. Kleinert, \pla{114} (1986) 263, \plb{174} (1987) 335.

\bibitem{my:inst} J.Pawe{\l}czyk, \plb{387} (1996) 287, hep-th/9607198.

\bibitem{my:holo}  J.Pawe{\l}czyk, \npb{491} (1997) 515, hep-th/9609196.

\bibitem{town} P.Howe, S.Penati, M.Pernici and P.Townsend, \plb{215} (1988)
555. 
\bibitem{ks}J.Kogut and L.Susskind, Phys.Rept. 8 (1973) 75.
\bibitem{glazek} Theory Of Hadrons And Light Front Qcd. Proceedings, 4th International
Workshop On Light Front Quantization And Nonperturbative Dynamics, Polana
Zgorzelisko, ed. S.D. Glazek, World Scientific, 1995. 

\bibitem{matrix} T. Banks, W. Fischler, S. H. Shenker, L. Susskind,
  Phys.Rev. D55 (1997) 5112, hep-th/9610043.
\bibitem{douglas} A.Connes, M.R. Douglas and A.Schwarz, Noncommutative Geometry and
Matrix Theory: Compactification on Tori, hep-th/9711162.  
\bibitem{witten:ym2} E.Witten, \jgp{9} (1992) 303.
 
\bibitem{hoppe} J.Hoppe, PhD thesis, MIT (1982);\\
E.G.Florates, J.Iliopoulos and G.Tiktopoulos, \plb{217} (1989) 285;\\
D.B.Fairlie and C.K.Zachos, \plb{224} (1989) 101;\\
C.N.Pope and K.S.Stelle,\plb{226} (1989) 257.

\bibitem{cmr}  S.Corder, G.Moore and S.Ramgoolam, 
\cmp{185} (1997) 543, hep-th/9402107.

\bibitem{horava}  P. Horava, 
Topological Strings and QCD in Two Dimensions, 
hep-th/9311156;\\ 
Nucl.Phys. B463 (1996) 238, hep-th/9507060.\\
A.M.Polyakov, Lectures given at the CRM-CAP Summer School { Particle and
Fields '94}, August 16-24 1994, Banff, Alberta,Canada.

\bibitem{gross} D. J. Gross, Nucl. Phys. B400 (1993) 161, hep-th/9212149
  ;D. J. Gross and W.Taylor, Nucl. Phys. B400 (1993) 
  181,hep-th/9301068; Nucl. Phys. B403 (1993) 395, hep-th/9303046.

\bibitem{sphere} M.R. Douglas and V.A.Kazakov, Phys. Lett. B319 (1993) 219, hep-th/9305047;
D.J. Gross and A. Matytsin, Nucl. Phys. B429 (1994) 50, hep-th/9404004.

\bibitem{mem} B. de Wit, J. Hoppe, H. Nicolai,
\npb{305} (1988) 545.

\bibitem{polch} S.Hellerman and J.Polchinski, Compactification in the Lightlike Limit,
hep-th/9711037.

\end{thebibliography}
\end{document}